\documentclass[twocolumn,trackchanges]{aastex62}
\usepackage{amsmath}
\hypersetup{
	colorlinks	= true,
	linkcolor	= red,
	urlcolor	= cyan,
	citecolor	= blue
}

\usepackage{amssymb}

\usepackage{lipsum}
\usepackage{multirow}

\graphicspath{{./images/}}

\newcommand{\exorelr}{\mbox{\textsc{ExoReL$^\Re$}}}

\turnoffediting

\shortauthors{Damiano \& Hu}

\usepackage{fancyhdr}
\begin{document}
	
	\title{Reflected spectroscopy of small exoplanets II: characterization of terrestrial exoplanets}
	
	\correspondingauthor{Mario Damiano}
	\email{mario.damiano@jpl.nasa.gov}
	
	\author[0000-0002-1830-8260]{Mario Damiano}
	\affiliation{Jet Propulsion Laboratory, California Institute of Technology, Pasadena, CA 91109, USA}
	
	\author[0000-0003-2215-8485]{Renyu Hu}
	\affiliation{Jet Propulsion Laboratory, California Institute of Technology, Pasadena, CA 91109, USA}
	\affiliation{Division of Geological and Planetary Sciences, California Institute of Technology, Pasadena, CA 91125, USA}
	
	\begin{abstract}
	A space telescope capable of high-contrast imaging has been recognized as the avenue toward finding terrestrial planets around nearby Sun-like stars and characterizing their potential habitability. It is thus essential to quantify the capability of reflected light spectroscopy obtained through direct imaging for terrestrial exoplanets, and existing work focused on planetary analogs of modern Earth. Here we go beyond Earth analogs and use a Bayesian retrieval algorithm, \exorelr, to determine what we could learn about terrestrial exoplanets from their reflected light spectra. Recognizing the potential diversity of terrestrial exoplanets, our focus is to distinguish atmospheric scenarios without any a priori knowledge of the dominant gas. We find that, while a moderate-resolution spectrum in the optical band ($0.4-1.0\ \mu$m) may sufficiently characterize a modern Earth analog, it would likely result in incorrect interpretation for planets similar to Archean Earth or having CO$_2$-dominated atmospheres. Including observations in the near-infrared bands ($1.0-1.8\ \mu$m) can prevent this error, determine the main component (N$_2$, O$_2$, or CO$_2$), and quantify trace gases (H$_2$O, O$_3$, and CH$_4$) of the atmosphere. These results are useful to define the science requirements and design the wavelength bandwidth and observation plans of exoplanet direct imaging missions in the future.
	\end{abstract}
	
	
	\section{Introduction} \label{sec:intro}
	
	To characterize the atmosphere, climate, and potential habitability of rocky exoplanets with spectroscopy is a significant endeavor that underpins the exploration of the Universe. The rocky exoplanets that have H$_2$-dominated atmospheres will be within the reach of the James Webb Space Telescope (JWST) for atmospheric studies through transits \citep[e.g.,][]{deming2009discovery,hu2021unveiling}. However, to use JWST to characterize moderate-temperate and rocky exoplanets that have N$_2$ or CO$_2$ atmospheres (i.e., potentially habitable planets that may resemble Earth) will likely require coadding a few tens of transits even for the very best targets due to the small vertical extent of the atmosphere \citep[e.g.,][]{belu2011primary,krissansen2018detectability,wunderlich2019detectability,pidhorodetska2020detectability,gialluca2021characterizing}.
	
	High-contrast imaging of exoplanets is a complementary and recognized avenue to enable the spectroscopic studies of temperate and rocky exoplanets. The level of starlight suppression required to image an Earth-like planet around a Sun-like star has been achieved in laboratories using both the coronagraph \citep[e.g.,][]{trauger2007laboratory} and starshade \citep[e.g.,][]{harness2021optical} technologies. The Nancy Grace Roman Space Telescope \citep[Roman,][]{Spergel2015,Akeson2019} will demonstrate the coronagraph technology in space. Studies of future large astrophysics missions commissioned by NASA have indicated that a sizable space telescope coupling with these starlight suppression technologies would be able to find small exoplanets in the habitable zones of nearby stars and study their atmospheres in the ultraviolet (UV), visible, and near-infrared (NIR) wavelengths \citep[][and HabEx Final Report\footnote{https://www.jpl.nasa.gov/habex/documents/} and LUVOIR Final Report\footnote{https://asd.gsfc.nasa.gov/luvoir/reports/}]{Roberge2018,Seager2019,Gaudi2020}. Based on these studies, the Astro2020 decadal survey recommends development toward a potential flagship mission to search for Earth-like habitable exoplanets through direct imaging \citep{national2021pathways}.
	
	Earth's spectrum in the UV, visible, and NIR wavelengths contain a wealth of information about the nature of its atmosphere \citep[e.g.,][]{Turnbull2006,kaltenegger2007spectral}. The spectrum comes from the reflected sunlight and has the absorption features of gases (e.g., H$_2$O, O$_2$, O$_3$) as well as clouds. \cite{Feng2018} applied an inverse retrieval algorithm to the spectrum of modern Earth -- as if it is imaged as an exoplanet -- to study to what extent one could learn about the atmospheric properties from the spectrum. With an N$_2$-dominated atmosphere assumed as the background atmosphere, \cite{Feng2018} demonstrated that the volume mixing ratios of H$_2$O, O$_2$, O$_3$ will be constrained from the spectrum. They have also explored a range of signal-to-noise ratios (S/N) and spectral resolutions (R), and found that to obtain meaningful constraints on the atmosphere, S/N should be at least 15 with a spectral resolution of R$=$140 and 10 if the spectral resolution is dropped to 70.
	
	Given the importance of the topic and the scarcity of studies, here we use our Bayesian retrieval method \citep[\exorelr,][]{Damiano2020a, Damiano2020b, Damiano2021} to explore the possible atmospheric constraints from reflected light spectra of terrestrial exoplanets. \exorelr \ preserves the causal relationship between a drop of the mixing ratio of H$_2$O and the resulting density of H$_2$O clouds, and thus typically yields constraints on the mixing ratio of H$_2$O below the clouds \citep{Damiano2020a, Damiano2021}. Also, in our previous work that focuses on the spectral retrieval of sub-Neptunes and water worlds \citep{Damiano2021}, we made key upgrades (i.e., the centered-log-ratio (CLR) formulation of gas mixing ratios and a new set of prior functions) that allowed us to correctly identify the dominant (i.e., background) gas in the atmosphere. This capability is important because sub-Neptunes may have a diverse atmospheric composition that ranges from H$_2$ to H$_2$O and CO$_2$ \citep[e.g.,][]{Zeng2019} and is also crucial for the current study.
	
	We use the retrieval algorithm to determine the main component of the atmosphere, rather than assuming the atmosphere to be N$_2$-dominated. This is important because we do not know whether terrestrial exoplanets to be found by direct-imaging missions will have N$_2$-dominated atmospheres. The rocky planets in the Solar System indicate that CO$_2$-dominated atmospheres are likely, and atmospheric formation models suggest a wide spectrum of plausible composition \citep[e.g.,][]{elkins2008ranges,gaillard2022redox}. An important objective of this study is to determine whether reflected light spectroscopy can distinguish types of rocky planets (e.g., Earth-like versus Venus-like) and characterize their atmospheric composition. We thus study the ability of spectral retrieval on atmospheric scenarios that correspond to not only modern Earth, but also Archean Earth \citep[e.g.,][]{catling2020archean} and planets with CO$_2$-dominated atmospheres that resemble Venus and Mars. This is, to our knowledge, the first time a retrieval method is applied to reflected light spectroscopy of small exoplanets beyond modern Earth analogs.
	
	Another important objective of the study is to determine the necessary wavelength coverage to detect and characterize the types of terrestrial exoplanets. We focus on the wavelength coverage, in addition to the S/N and spectral resolution as the observational requirements \citep[e.g., ][]{Feng2018}, because it is one of the most important factors that impact the overall architecture of future exoplanet imaging missions. For example, the coronagraph instruments of future missions would probably operate with a wavelength bandpass of $\sim20\%$ \citep[e.g.,][]{Roberge2018}, and so to obtain a spectrum from 0.4 to, for example, 1.8 $\mu$m, $\sim8$ integrations would be necessary -- a coverage to only 1.0 $\mu$m would halve this requirement. The starshade could result in a wider wavelength bandpass of $\sim100\%$ achievable in one shot \citep{Gaudi2020}, but the starshade must be moved to different separations from the telescope to operate between the optical ($0.4-1.0\ \mu$m) and the NIR band ($1.0-1.8\ \mu$m), which consumes fuels. Therefore, it is essential to assess whether obtaining the spectrum in the NIR band would be necessary to sufficiently characterize the atmospheres of terrestrial exoplanets.
	
	The paper is organized as follows: Sec. \ref{sec:model} describes the setup of the retrieval algorithm and the simulation of the atmospheric scenarios, Sec. \ref{sec:result} shows the results of the retrievals and compares the posterior constraints at varied wavelength coverage, spectral resolution, and signal-to-noise ratio for the Earth-like scenario and other atmospheric scenarios. In Sec. \ref{sec:discussion}, we discuss the results and the implications on the characterization of terrestrial exoplanets through reflected light spectroscopy and the ramification on the design of future missions. Finally, Sec. \ref{sec:conclusion} summarizes the key findings of this study.
	
	\section{Methods} \label{sec:model}
	
	\subsection{Retrieval Setup} \label{sec:retrieval}
	
	As described in \cite{Damiano2021}, we have updated the radiative transfer model of \exorelr\ to enable simulations of the reflected light spectra of small exoplanets.  We have implemented the centered-log-ratio \citep{Aitchison1982} of the mixing ratios of H$_2$O, CO$_2$, O$_2$, O$_3$, and CH$_4$ as free parameters, and N$_2$ as the filler gas, and applied the set of priors proposed in \cite{Damiano2021} to ensure that no one gas, including the filler gas, would be preferred as the background gas a priori.
	
\begin{figure}
	    \centering
	    \includegraphics[scale=0.6]{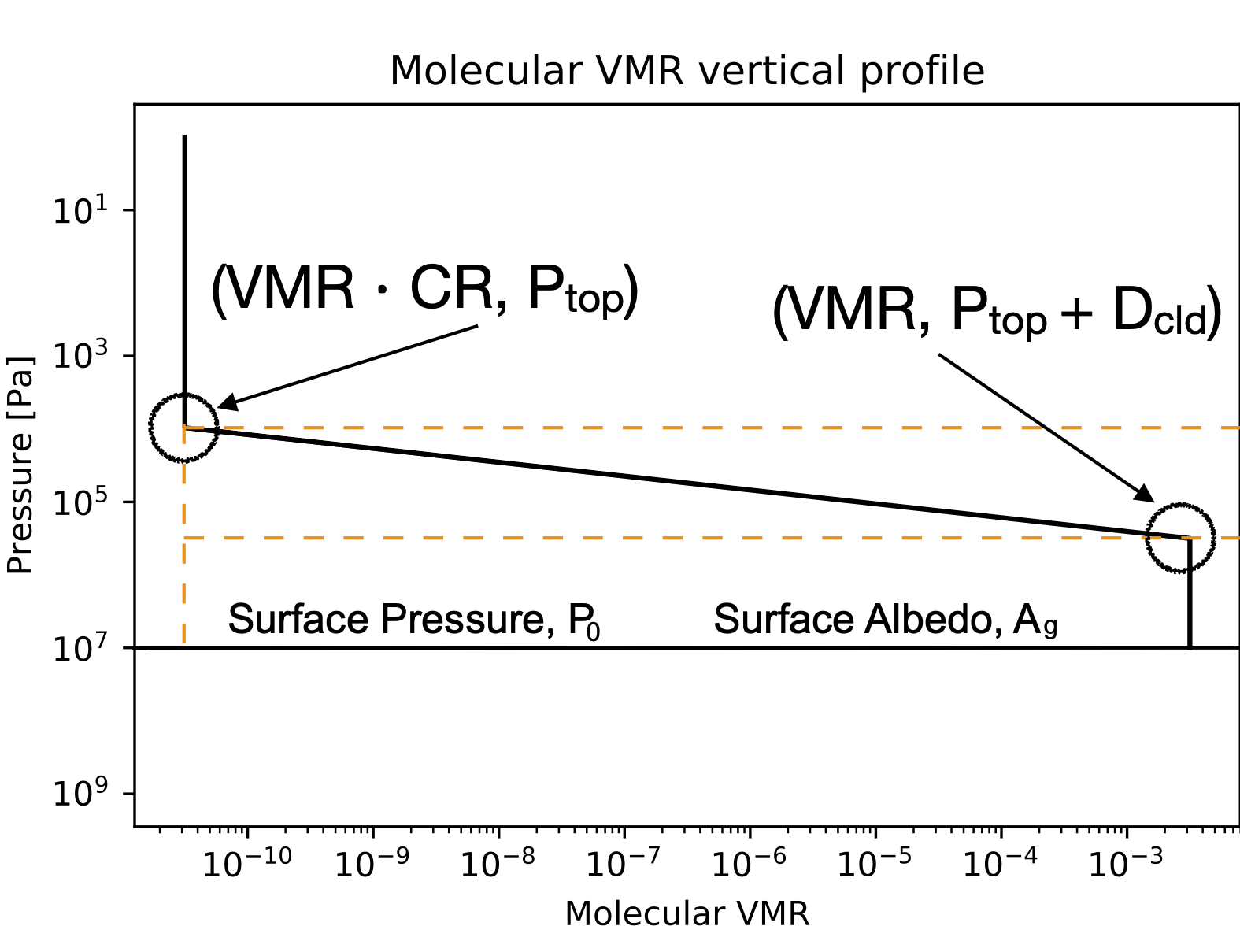}
	    \caption{Parameterization of the volume mixing ratio (VMR) vertical profile of the condensing gas (i.e., water vapor) in this study. At the bottom of the atmosphere, we include a reflecting surface characterized by two free parameters: the surface pressure and the surface albedo. The non-constant VMR allow for the presence of clouds where the gas condenses (i.e., decreases the gas VMR). The four parameters that define the VMR profile and thus the cloud density profile highlighted in the schematics are also free parameters of the model.}
	    \label{fig:profile}
	\end{figure}

\begin{figure}[!h]
	\centering
	\includegraphics[width=0.45\textwidth]{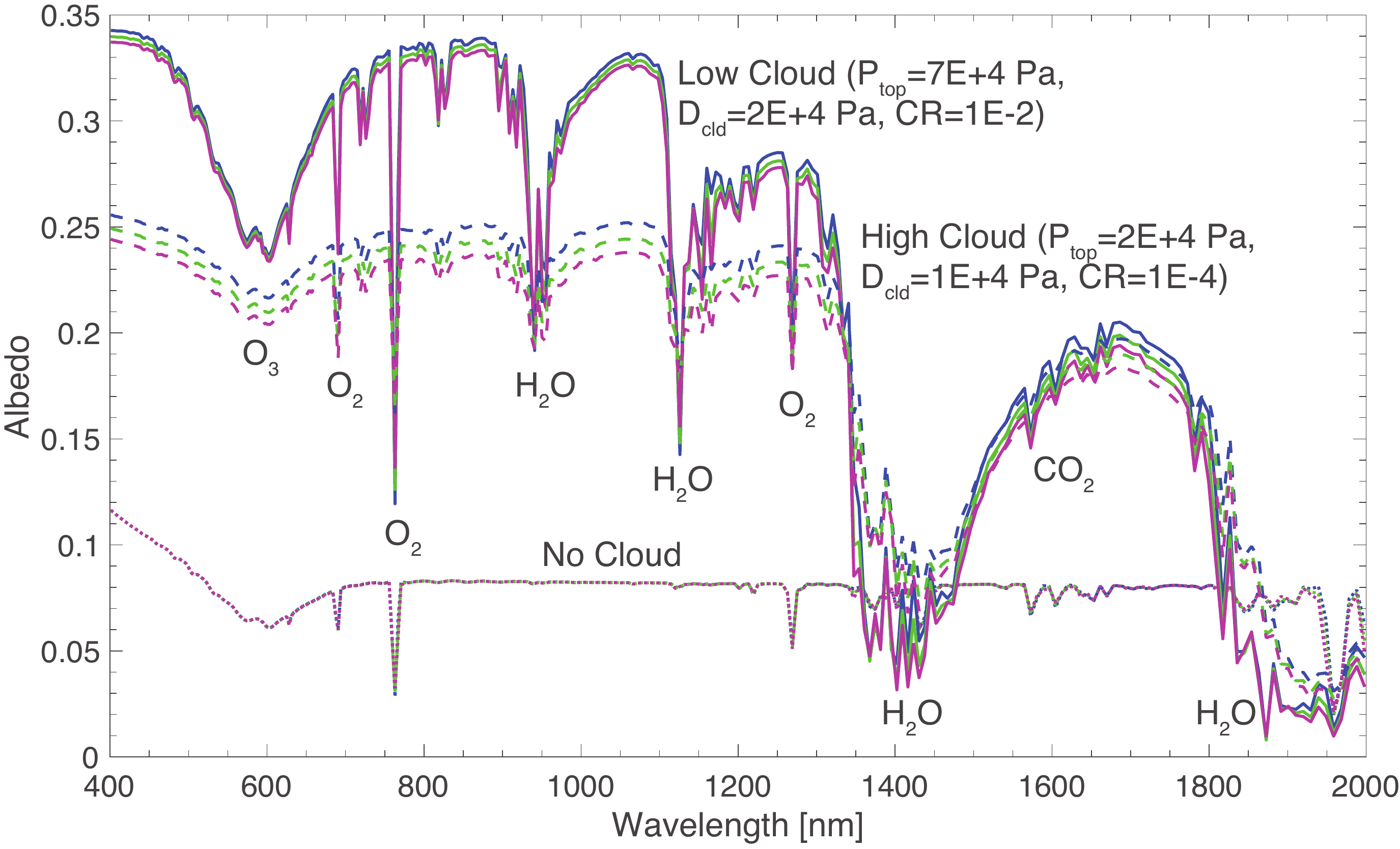}
	\caption{\label{fig:earth}Modeled spectra of an atmosphere that has a surface pressure of $10^5$ Pa (1 bar) and composition similar to modern Earth's atmosphere. Solid, dashed, and dotted lines correspond to low cloud, high cloud, no cloud scenarios, and the three colors within each group correspond to an assumed temperature of 200, 250, and 300 K. The orbital phase is $\pi/3$. The surface albedo is 0.2. The spectra are not sensitive to the assumed temperature.}
	\end{figure}

	A main upgrade of the model is to include the possibility to detect a planetary surface through the characterization of light reflected by a rocky surface. The current model additionally includes the surface pressure and surface albedo as free parameters (see Fig. \ref{fig:profile}). We assume the surface to be isotropically scattering, with the surface albedo as a free parameter. Similar to \cite{Feng2018}, we do not include the wavelength dependency of the surface albedo \citep{hu2012theoretical}. This aspect will be addressed by a future study. Also, instead of calculating the pressure-temperature profile, we assume the atmosphere to be isothermal and assign an estimated temperature given the irradiation level. This greatly simplifies the forward model and is valid because the reflected light spectra are not sensitive to the specifics of the temperature \citep[][also see Fig. \ref{fig:earth}]{Feng2018}.
	
	In this study, we include water clouds as the only type of condensates in the atmosphere. This is consistent with the current plans for direct imaging missions. We use the optical properties from \cite{Palmer1974} to calculate the cross-sections and single scattering albedo of water droplets. In this way, our model includes the absorption bands of water clouds at 1.44 $\mu$m and 1.93 $\mu$m, and gradually more absorption of the clouds outside these bands when the wavelength is longer than $\sim1$ $\mu$m. The description of the cloud follows \cite{Damiano2020a}, which preserves the causal relationship between the condensation of water and the formation of the cloud (Fig.~\ref{fig:profile}). In our model, the cloud is parameterized by the cloud top pressure ($P_{\rm top}$), the cloud depth (i.e., the difference in pressure between the cloud bottom and the cloud top, $D_{\rm cld}$), and the condensation ratio (i.e., the ratio between the mixing ratio of water above the cloud and that below the cloud, denoted as CR).
	
	Similar to the test cases presented in Fig. 1 of \cite{Feng2018}, we have run test cases at the limit of no atmospheric absorption or scattering and confirm that the phase function of the planet to be Lambertian, scaled by the assumed surface albedo.
	
	For a second test, we model the albedo spectra of an Earth-like planet having varied cloud scenarios that correspond to Earth's cumulus clouds and cirrus clouds (Fig. \ref{fig:earth}). The albedo spectra are highly sensitive to the cloud scenario and are not sensitive to the temperature. The subtle difference in the spectra due to temperature is caused by the cloud particle size's mild dependency on temperature, which is described in \cite{Hu2019B2019ApJ...887..166H}. Absorption features of O$_2$, O$_3$, H$_2$O, and CO$_2$ are visible, and the descending slope towards longer wavelengths is a combined effect of stronger water absorption and single scattering albedo of cloud particles. The combination of the three components shown in Fig. \ref{fig:earth} gives a spectrum that resembles Earth's reflected light spectrum observed in Earthshine \citep{Turnbull2006}.
	
	\subsection{Simulated Atmospheric Scenarios} \label{sec:scenarios}
	
	Focusing on the atmospheric characterization of terrestrial exoplanets, we simulate a 1M$_{\oplus}$, 1R$_{\oplus}$ planet at 1AU from a Sun-like star. We surround the planet with different atmospheric scenarios to explore the capability of reflected light spectroscopy.
	
	We have studied four atmospheric scenarios in this work. (1) An Earth-like atmosphere. Similar to \citep{Feng2018}, we also consider a rocky planet having an atmosphere skin to modern Earth's atmosphere, which is an N$_2$-dominated atmosphere ($\sim$ 79\%) with the presence of O$_2$ ($\sim$ 20\%) and other minor gases including H$_2$O, O$_3$, CH$_4$, and CO$_2$ (Table~\ref{tab:earthlike}). (2) An Archean Earth-like atmosphere. Earth's atmospheres has gone through multiple evolution stages and had a very different composition in the past \citep[e.g.,][]{catling2020archean}. The paleoatmosphere during the Archean eon is particularly interesting because Archean corresponds to $>30\%$ of Earth's history and the atmosphere at the time had a low mixing ratio of O$_2$ but much more CO$_2$ and CH$_4$ than the present day (Table~\ref{tab:archean}). (3) A CO$_2$-dominated atmosphere with O$_2$ and clouds. In this scenario we start with modern Earth's atmosphere but swap the mixing ratios between N$_2$ and CO$_2$, effectively creating a hypothetical CO$_2$-dominated atmosphere with O$_2$ (Table~\ref{tab:co2clouds}). The motivation of this scenario is two-fold. First, we want to test whether the reflected light spectrum would tell us that the planet should have a CO$_2$-dominated atmosphere, simulating the detection of a Venus-like planet. Second, we want to study whether O$_2$, a main biosignature gas \citep[e.g.,][]{schwieterman2018exoplanet}, would still be detectable in such an atmosphere. (4) A dry CO$_2$-dominated atmosphere. The last scenario we simulate is an arid planet with a CO$_2$-dominated atmosphere ($\sim$ 99\%). As an endmember, we consider the case without any water vapor in the atmosphere, and thus no water clouds either (Table~\ref{tab:co2sky}). This scenario may correspond to a rocky planet that has lost its ocean \citep[e.g.,][]{wordsworth2013water}, or a rocky planet that is too cold for water to exist in liquid or vapor forms with appreciable amounts (such as Mars). We consider minor presence of CH$_4$ and O$_2$ to assess the detectability of these gases in such an atmosphere.
	
	In all scenarios, we study the impact of including and excluding the NIR portion of the reflected light spectra in the retrieval. In this work, we adopt the ``optical'' wavelengths to $0.4-1.0\ \mu$m and the ``NIR'' wavelengths to $1.0-1.8\ \mu$m, and adopt a baseline spectral resolution of R$=$140 and R$=$40 for the optical and NIR bands, respectively. These assumptions are approximately consistent with the large mission studies \citep{Roberge2018,Gaudi2020}. Additionally for the Earth-like scenario, we explore a signal-to-noise ratio (S/N) of 5, 10, and 20, as well as the spectral resolution in the optical band between 70 and 140. This exploration allows us to compare with previous work \citep{Feng2018} and better understand the observational capabilities required to extract meaningful information from the spectrum. 
	
	Lastly, to simulate the size of the error associated to each of the data points, we considered the maximum value of the spectrum and we divided it by the desired S/N. This effectively assumes that the noise is dominated by the background and not the planet, likely valid for spectroscopy of terrestrial planets \citep[e.g.,][]{hu2021overview}. We then added a Gaussian deviation to the data to simulate the random realization of the observation.
	
	\begin{deluxetable*}{cccccc}
		\tablecaption{Atmospheric parameters used to simulate the Earth-like scenario and the retrieval results. The spectral resolution R is applied to the optical band of 0.4 -- 1.0 $\mu$m, and the spectrum in the NIR band of 1.0 -- 1.8 $\mu$m with R$=$40 and the same S/N as listed in the table is also included in the retrieval. The error bars of the retrieval results correspond to the 95\% confidence interval (i.e., 2$\sigma$). Note -- the column of S/N$=$10, R$=$140 reports the weighted average of the two degenerate solutions found by the retrieval. \label{tab:earthlike}}
		\tablehead{
			\colhead{Parameter} & \colhead{Input} & \colhead{S/N$=$5, R$=$70} & \colhead{S/N$=$10, R$=$70} & \colhead{S/N$=$10, R$=$140} & \colhead{S/N$=$20, R$=$140}}
		\startdata
		$Log(P_{0})$ [Pa] & $5.00$ & $7.25^{+3.39}_{-3.08}$ & $7.35^{+3.22}_{-2.65}$ & $8.00^{+2.70}_{-3.23}$ & $6.84^{+3.47}_{-1.98}$ \\
		$Log(P_{top, H_2O})$ [Pa] & $4.85$ & $1.50^{+2.58}_{-1.39}$ & $4.03^{+0.80}_{-2.23}$ & $3.96^{+3.08}_{-1.06}$ & $4.26^{+0.50}_{-0.41}$ \\
		$Log(D_{cld, H_2O})$ [Pa] & $4.30$ & $4.97^{+1.59}_{-1.73}$ & $4.42^{+0.44}_{-0.77}$ & $4.34^{+2.56}_{-0.39}$ & $4.48^{+0.28}_{-0.26}$ \\
		$Log(CR_{H_2O})$ & $-3.00$ & $-4.66^{+3.67}_{-6.52}$ & $-8.45^{+5.25}_{-3.26}$ & $-8.30^{+5.02}_{-3.36}$ & $-8.21^{+4.75}_{-3.27}$ \\
		$Log(VMR_{H_2O})$ & $-2.01$ & $-0.08^{+0.08}_{-1.22}$ & $-1.29^{+1.23}_{-0.62}$ & $-1.46^{+0.89}_{-2.47}$ & $-1.69^{+0.44}_{-0.47}$ \\
		$Log(VMR_{CH_4})$ & $-5.96$ & $-6.41^{+2.70}_{-3.62}$ & $-6.97^{+2.38}_{-3.29}$ & $-7.32^{+2.60}_{-3.38}$ & $-5.22^{+0.73}_{-3.71}$ \\
		$Log(VMR_{CO_2})$ & $-3.40$ & $-3.81^{+3.54}_{-5.95}$ & $-3.23^{+2.13}_{-5.42}$ & $-5.53^{+3.56}_{-4.92}$ & $-4.36^{+1.97}_{-4.14}$ \\
		$Log(VMR_{O_2})$ & $-0.71$ & $-2.39^{+2.31}_{-6.47}$ & $-0.03^{+0.03}_{-1.23}$ & $-0.04^{+0.03}_{-4.18}$ & $-0.34^{+0.23}_{-0.59}$ \\
		$Log(VMR_{O_3})$ & $-5.96$ & $-4.99^{+0.91}_{-0.95}$ & $-5.59^{+0.81}_{-0.47}$ & $-5.43^{+0.25}_{-2.44}$ & $-5.66^{+0.17}_{-0.36}$ \\
		$Log(VMR_{N_2})$ & $-0.10$ & $-2.63^{+2.59}_{-5.98}$ & $-3.84^{+3.40}_{-6.47}$ & $-1.90^{+1.90}_{-4.86}$ & $-0.29^{+0.23}_{-0.45}$ \\
		$A_g$ & $0.05$ & $0.53^{+0.42}_{-0.47}$ & $0.53^{+0.42}_{-0.47}$ & $0.46^{+0.48}_{-0.42}$ & $0.40^{+0.50}_{-0.35}$ \\
		$Log(g\ [cgs])$ & $2.99$ & $3.00^{+0.04}_{-0.04}$ & $3.01^{+0.03}_{-0.03}$ & $3.00^{+0.03}_{-0.04}$ & $2.99^{+0.02}_{-0.01}$ \\
		$\mu$ & $28.70$ & $20.46^{+13.09}_{-2.44}$ & $31.26^{+0.98}_{-11.40}$ & $30.94^{+0.79}_{-4.25}$ & $29.62^{+1.24}_{-1.31}$ \\
		\enddata
	\end{deluxetable*}

	\section{Results} \label{sec:result}
	
	\subsection{Earth-like scenario} 
	\label{sec:earth-like}
	
\begin{figure*}[!h]
		\centering
		\includegraphics[scale=0.45]{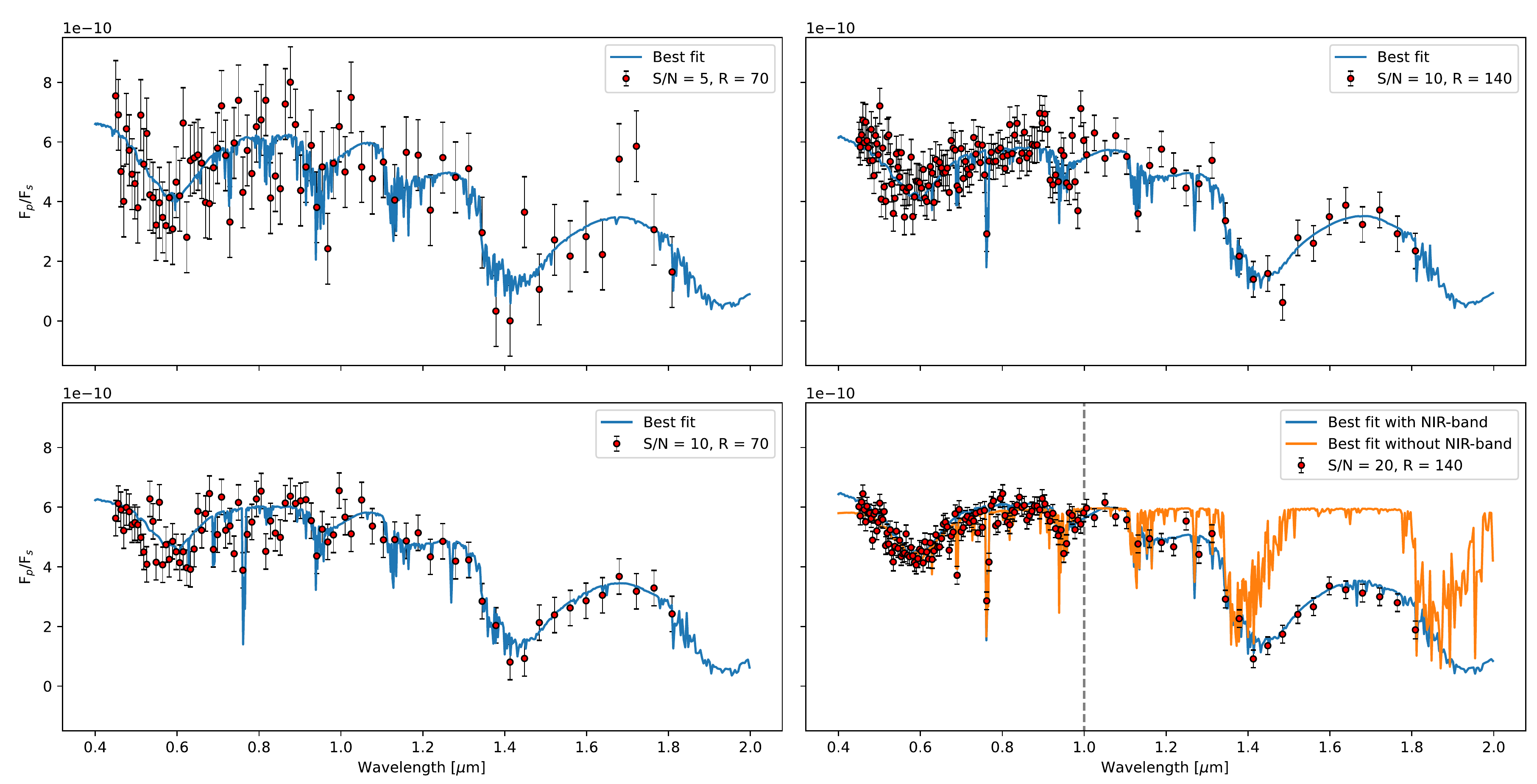}
		\caption{The simulated data and the best-fit spectra of the Earth-like scenario. Combinations of different S/N and optical-band spectral resolutions have been considered to generate the four different cases. The blue lines depict the best-fit spectra from the retrieval, and the orange line depicts the best-fit spectrum without fitting the NIR band. \label{fig:earth-like}}
	\end{figure*}

	The retrieval results are reported in Table~\ref{tab:earthlike} and the posterior distributions are shown in Appendix~\ref{sec:A_earth}. The simulated data and the best-fit spectra are shown in Fig.~\ref{fig:earth-like}. 
    
    The retrieval algorithm is able to successfully determine the basic nature of the planet (i.e., having an N$_2$/O$_2$-dominated atmosphere with H$_2$O and O$_3$) with the optical and the NIR bands, except for the case of S/N$=$5 and R$=$70. In this minimal-quality case, the noise realization dominates the spectral features and the retrieval algorithm mistakenly returns a scenario closer to a water world (i.e., having an H$_2$O-dominated atmosphere). For the other cases, the algorithm returns a fit that corresponds to the input atmospheric composition. The abundance of H$_2$O and O$_3$ is generally well constrained, but that of CH$_4$ is only moderately constrained in the best-case scenario with S/N$=$20 and R$=$140. The abundance of CO$_2$ is not well constrained in any of the scenarios. The cloud pressures are reasonably well constrained, but the surface pressure (except that it must be greater than the cloud pressure) or the surface albedo cannot be constrained. We see that the constraints of these parameters are the tightest in the case of S/N$=$20 and R$=$140, as expected.
    
    Interestingly, the case with S/N$=$10 and R$=$140 (i.e., high resolution with poor S/N) results in a degenerate solution that involves lower clouds and a lower mixing ratio of O$_2$ than the truth. While this degenerate solution may be deemed unphysical (e.g., the retrieval still indicates a significant mixing ratio of O$_3$), this finding suggests that it may be advisable to test the sensitivity of the retrieval results on multiple choices of spectral binning. 
    
    We use the case R$=$140 and S/N$=$20 to explore the effect of the NIR part of the spectrum (1.0 -- 1.8 $\mu$m) on the atmospheric constraints. The retrieval results with and without the NIR band are compared in Table~\ref{tab:earthlikeIR}, and the posterior distributions are compared in Appendix~\ref{sec:A_earth}. Without the NIR band but with the ``high-quality'' option (S/N$=$20 and R$=$140) for the optical band, the retrieval algorithm returns a solution that is moderately different from the truth -- an O$_2$-dominated atmosphere, with large biases in the VMR of H$_2$O and the cloud properties. The constraints on the VMR of O$_3$ remain solid. The degradation in the retrieval quality ultimately comes from the fact that the NIR band contains information about the pressure of clouds, as the water droplet is not fully reflective in the NIR band and can cause an underlying slope outside of the gas absorption bands (see Fig.~\ref{fig:earth}). From a bigger-picture standpoint, we feel that this interpretation can be deemed ``somewhat correct'' since it still calls for an O$_2$/O$_3$-rich atmosphere with H$_2$O.
    
    \begin{deluxetable}{cccc}
		\tablecaption{Atmospheric parameters used to simulate the Earth-like scenario and retrieval results for the full wavelength band and only the optical band. The error bars of the retrieval results correspond to the 95\% confidence interval (i.e., 2$\sigma$). \label{tab:earthlikeIR}}
		\tablehead{
			\colhead{Parameter} & \colhead{Input} & \colhead{Optical $+$ NIR} & \colhead{Optical only}}
		\startdata
		$Log(P_{0})$ [Pa] & $5.00$ & $6.84^{+3.47}_{-1.98}$ & $4.39^{+0.13}_{-0.12}$\\
		$Log(P_{top, H_2O})$ [Pa] & $4.85$ & $4.26^{+0.50}_{-0.41}$ & $2.07^{+4.60}_{-1.89}$\\
		$Log(D_{cld, H_2O})$ [Pa] & $4.30$ & $4.48^{+0.28}_{-0.26}$ & $1.79^{+4.43}_{-1.63}$\\
		$Log(CR_{H_2O})$ & $-3.00$ & $-8.21^{+4.75}_{-3.27}$ & $-6.57^{+6.07}_{-4.93}$\\
		$Log(VMR_{H_2O})$ & $-2.01$ & $-1.69^{+0.44}_{-0.47}$ & $-3.21^{+1.76}_{-0.29}$\\
		$Log(VMR_{CH_4})$ & $-5.96$ & $-5.22^{+0.73}_{-3.71}$ & $-6.34^{+1.85}_{-2.83}$\\
		$Log(VMR_{CO_2})$ & $-3.40$ & $-4.36^{+1.97}_{-4.14}$ & $-3.39^{+2.25}_{-4.00}$\\
		$Log(VMR_{O_2})$ & $-0.71$ & $-0.34^{+0.23}_{-0.59}$ & $-0.00^{+0.00}_{-0.05}$\\
		$Log(VMR_{O_3})$ & $-5.96$ & $-5.66^{+0.17}_{-0.36}$ & $-5.52^{+0.14}_{-0.14}$\\
		$Log(VMR_{N_2})$ & $-0.10$ & $-0.29^{+0.23}_{-0.45}$ & $-3.65^{+2.58}_{-4.64}$\\
		$A_g$ & $0.05$ & $0.40^{+0.50}_{-0.35}$ & $0.64^{+0.31}_{-0.24}$\\
		$Log(g\ [cgs])$ & $2.99$ & $2.99^{+0.02}_{-0.01}$ & $2.89^{+0.17}_{-0.20}$\\
		$\mu$ & $28.70$ & $29.62^{+1.24}_{-1.31}$ & $31.99^{+0.83}_{-0.63}$\\
		\enddata
	\end{deluxetable}
    
    \subsection{Archean Earth-like scenario}
    
    The simulated data and the best-fit models calculated by \exorelr\ are shown in Fig.~\ref{fig:S1_e}, and the full posterior distributions are reported in Appendix~\ref{sec:A_archean}. The synthesized spectrum shows more deep absorption features in the red end of the optical and the NIR band compared to the Earth-like scenario, and this is due to the higher mixing ratios of CH$_4$ and CO$_2$ in the atmosphere.
	
	The retrieval algorithm correctly characterizes the atmosphere by determining that N$_2$ is the dominant gas and that the atmosphere has high mixing ratios of CO$_2$ and CH$_4$, with the full spectrum that includes both the optical and NIR bands. The retrieval algorithm also correctly determines that the atmosphere has low amounts of O$_2$ and O$_3$. The pressure of the water cloud, as well as the mixing ratio of water vapor below the cloud, have also been correctly retrieved within 1$\sigma$.
	
	\begin{figure}[!h]
		\centering
		\includegraphics[width=\columnwidth]{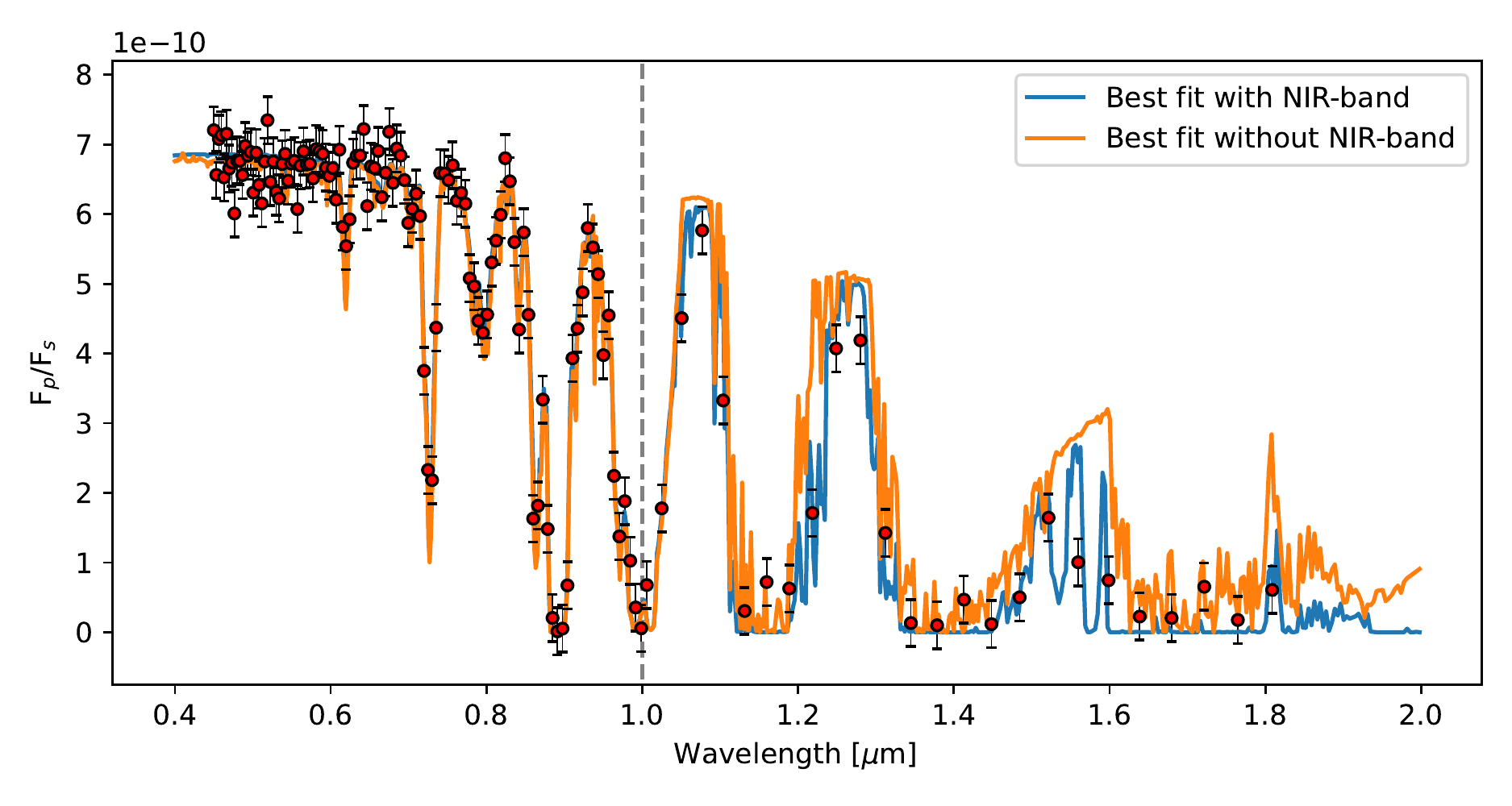}
		\caption{The simulated data of the Archean Earth-like scenario (red points) and the best-fit models with (blue line) and without the NIR band (orange line). \label{fig:S1_e}}
	\end{figure}
	
	\begin{deluxetable}{cccc}
		\tablecaption{Atmospheric parameters used to simulate the Archean Earth-like scenario and the retrieval results. The error bars of the retrieval results correspond to the 95\% confidence interval (i.e., 2$\sigma$). \label{tab:archean}}
		\tablehead{
			\colhead{Parameter} & \colhead{Input} & \colhead{Optical $+$ NIR} & \colhead{Optical only}}
		\startdata
		$Log(P_{0})$ [Pa] & $5.00$ & $7.39^{+3.19}_{-2.21}$ & $7.81^{+2.75}_{-3.68}$\\
		$Log(P_{top, H_2O})$ [Pa] & $4.85$ & $4.79^{+0.63}_{-0.33}$ & $2.28^{+1.28}_{-1.58}$\\
		$Log(D_{cld, H_2O})$ [Pa] & $4.30$ & $4.58^{+0.52}_{-0.33}$ & $3.51^{+0.48}_{-0.27}$\\
		$Log(CR_{H_2O})$ & $-4.00$ & $-9.00^{+4.24}_{-2.73}$ & $-9.08^{+4.43}_{-2.64}$\\
		$Log(VMR_{H_2O})$ & $-1.00$ & $-0.72^{+0.43}_{-0.69}$ & $-0.12^{+0.09}_{-0.14}$\\
		$Log(VMR_{CH_4})$ & $-2.00$ & $-2.03^{+0.30}_{-0.60}$ & $-0.67^{+0.26}_{-0.59}$\\
		$Log(VMR_{CO_2})$ & $-1.00$ & $-0.71^{+0.40}_{-0.97}$ & $-4.94^{+3.47}_{-4.41}$\\
		$Log(VMR_{O_2})$ & $-7.00$ & $-5.30^{+2.15}_{-2.81}$ & $-4.43^{+2.65}_{-4.42}$\\
		$Log(VMR_{O_3})$ & $-$ & $-8.09^{+0.82}_{-1.13}$ & $-7.21^{+1.22}_{-1.88}$\\
		$Log(VMR_{N_2})$ & $-0.10$ & $-0.25^{+0.20}_{-0.32}$ & $-2.62^{+1.90}_{-3.41}$\\
		$A_g$ & $0.05$ & $0.64^{+0.33}_{-0.55}$ & $0.45^{+0.47}_{-0.40}$\\
		$Log(g\ [cgs])$ & $2.99$ & $2.99^{+0.01}_{-0.01}$ & $2.98^{+0.01}_{-0.02}$\\
		$\mu$ & $28.49$ & $28.98^{+5.33}_{-4.60}$ & $17.75^{+2.30}_{-0.46}$\\
		\enddata
	\end{deluxetable}
	
	Without the NIR band, the nature of the atmosphere is not correctly characterized. In this case (see Appendix \ref{sec:A_archean}), \exorelr\ reports an H$_2$O-dominated atmosphere with a high cloud (at pressures $<\sim0.01$ bar). Because of this, the retrieved mixing ratio of CH$_4$ is biased toward larger values. The retrieval algorithm cannot detect CO$_2$, but rather returns an erroneous upper limit of $\sim1\%$. Evidently, the retrieval algorithm tends to misinterpret the spectrum in the Archean Earth case without the NIR band. This is because the absorption features of gaseous CO$_2$ and the clouds of H$_2$O are both in the NIR band -- without these ``anchors'' the retrieval algorithm would return a wrong solution.
    
    \subsection{CO$_2$-dominated atmosphere with O$_2$ and clouds}
    
    The simulated data and the best-fit models calculated by \exorelr\ are shown in Fig.~\ref{fig:S1_d}, and the full posterior distribution is reported in Appendix~\ref{sec:A_co2}. The spectrum is similar to the Earth-like scenario in the optical band but has a different shape in the NIR band mainly due to strong CO$_2$ absorption. 
    
    The retrieval algorithm is able to correctly characterize the atmosphere of the simulated planet from the spectrum that includes the optical and the NIR bands. The posteriors clearly show a CO$_2$-dominated atmosphere with O$_2$ as the second most dominant gas. The abundance of H$_2$O and O$_3$ and the cloud pressures are well constrained, and the abundance of CH$_4$ is poorly (but correctly) constrained. The retrieval algorithm indicates that the data can tolerate an abundance of N$_2$ (a spectrally inactive gas) that is much higher than the input abundance, and thus the calculated mean molecular mass $\mu$ is lower than the expected value. But this does not affect the overall interpretation of the nature of the atmosphere.
    
    \begin{figure}[!h]
		\centering
		\includegraphics[width=\columnwidth]{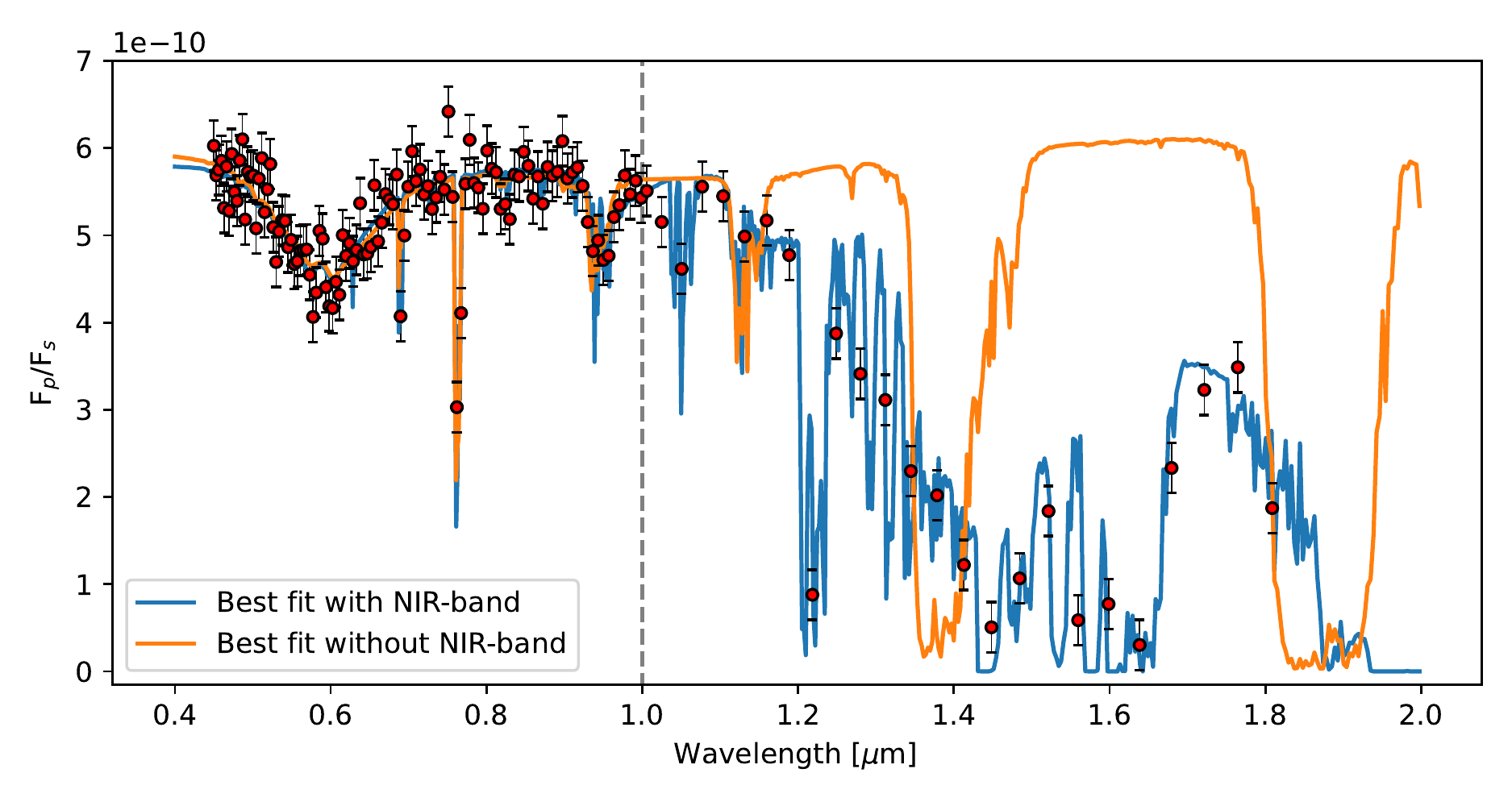}
		\caption{The simulated data of the CO$_2$-dominated atmosphere with O$_2$ and clouds scenario (red points) and the best-fit models with (blue line) and without the NIR band (orange line). \label{fig:S1_d}}
	\end{figure}
	
	\begin{deluxetable}{cccc}
		\tablecaption{Atmospheric parameters used to simulate the CO$_2$-dominated atmosphere with O$_2$ and clouds scenario and the retrieval results. The error bars of the retrieval results correspond to the 95\% confidence interval (i.e., 2$\sigma$). \label{tab:co2clouds}}
		\tablehead{
			\colhead{Parameter} & \colhead{Input} & \colhead{Optical $+$ NIR} & \colhead{Optical only}}
		\startdata
		$Log(P_{0})$ [Pa] & $5.00$ & $5.80^{+2.77}_{-0.60}$ & $6.98^{+0.23}_{-0.24}$\\
		$Log(P_{top, H_2O})$ [Pa] & $4.85$ & $4.88^{+0.25}_{-0.26}$ & $3.05^{+2.98}_{-2.83}$\\
		$Log(D_{cld, H_2O})$ [Pa] & $4.30$ & $4.84^{+0.19}_{-0.27}$ & $3.11^{+2.97}_{-2.87}$\\
		$Log(CR_{H_2O})$ & $-3.00$ & $-8.33^{+4.37}_{-3.29}$ & $-5.46^{+5.06}_{-6.03}$\\
		$Log(VMR_{H_2O})$ & $-2.01$ & $-2.14^{+0.65}_{-0.50}$ & $-6.39^{+0.27}_{-0.28}$\\
		$Log(VMR_{CH_4})$ & $-5.96$ & $-7.07^{+1.40}_{-1.52}$ & $-8.71^{+1.75}_{-2.18}$\\
		$Log(VMR_{CO_2})$ & $-0.10$ & $-0.32^{+0.15}_{-0.34}$ & $-7.12^{+4.64}_{-3.60}$\\
		$Log(VMR_{O_2})$ & $-0.71$ & $-1.16^{+0.25}_{-0.46}$ & $-3.83^{+0.23}_{-0.25}$\\
		$Log(VMR_{O_3})$ & $-5.96$ & $-6.21^{+0.12}_{-0.22}$ & $-8.01^{+0.20}_{-0.19}$\\
		$Log(VMR_{N_2})$ & $-3.40$ & $-0.35^{+0.23}_{-0.30}$ & $-0.00^{+0.00}_{-0.00}$\\
		$A_g$ & $0.05$ & $0.36^{+0.52}_{-0.29}$ & $0.96^{+0.03}_{-0.04}$\\
		$Log(g\ [cgs])$ & $2.99$ & $2.99^{+0.02}_{-0.03}$ & $3.04^{+0.01}_{-0.02}$\\
		$\mu$ & $43.54$ & $35.78^{+3.12}_{-4.26}$ & $28.01^{+0.05}_{-0.00}$\\
		\enddata
	\end{deluxetable}
	
	Without the NIR band, however, the retrieval algorithm significantly misinterprets the nature of the atmosphere. In this case, the retrieval algorithm mistakenly reports a 100-bar N$_2$-dominated atmosphere, without clouds, and with mixing ratios of H$_2$O, O$_2$, and O$_3$ smaller than the true values by several orders of magnitudes. The retrieval algorithm also mistakenly reports an upper limit of CO$_2$ of approximately 1\% and suggested the planet would have a highly reflective surface. Evidently the spectral interpretation without the NIR band would be wrong, and this is again because the absorption features of gaseous CO$_2$ and the clouds of H$_2$O are both in the NIR band.
	
	\subsection{Dry CO$_2$-dominated atmosphere}
	
\begin{figure}[!h]
		\centering
		\includegraphics[width=\columnwidth]{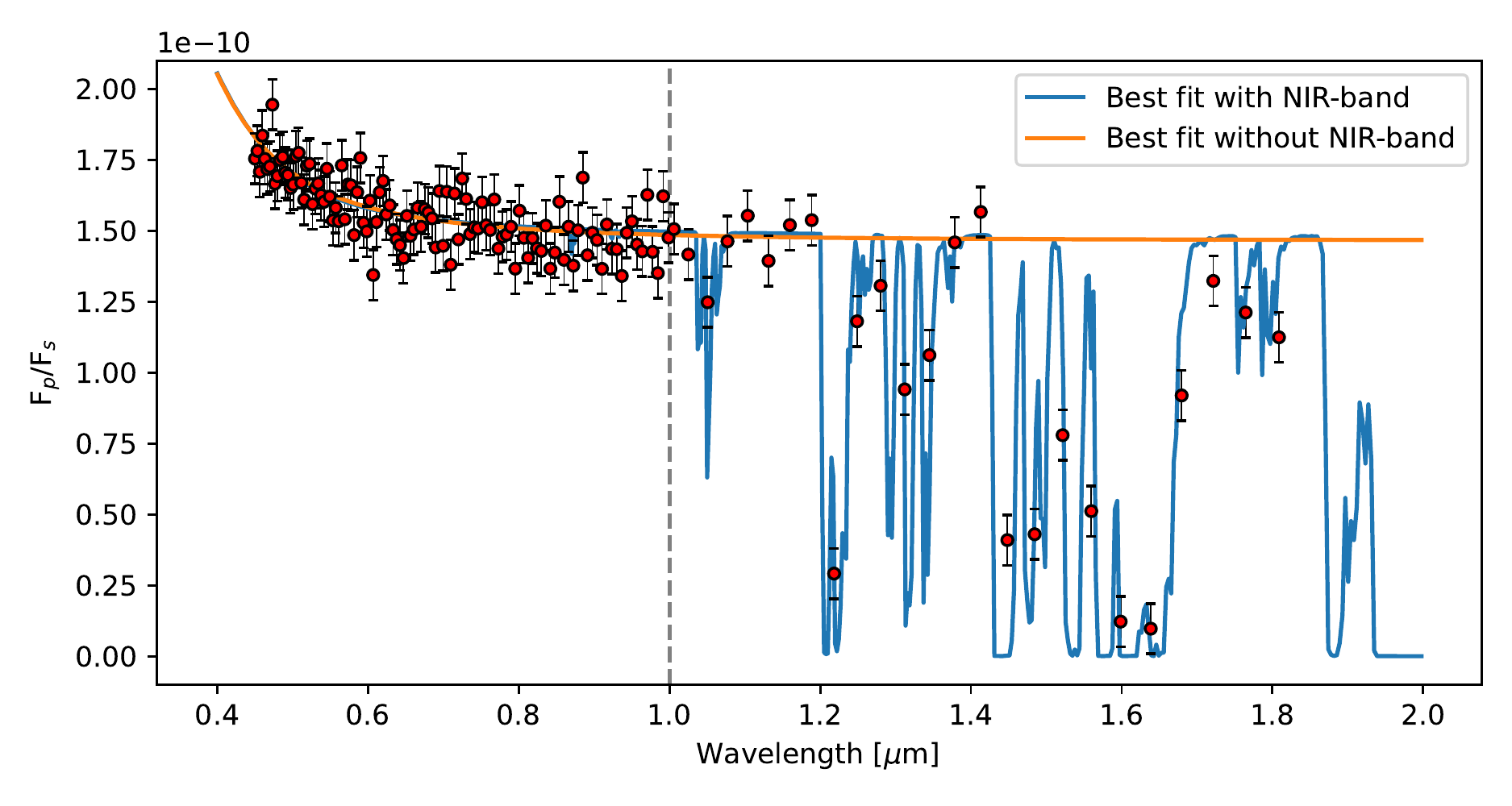}
		\caption{The simulated data of the dry CO$_2$-dominated atmosphere scenario (red points) and the best-fit models with (blue line) and without the NIR band (orange line). \label{fig:S1_c}}
	\end{figure}

	The simulated data and the best-fit models calculated by \exorelr\ are shown in Fig.~\ref{fig:S1_c}. The spectrum on the blue end is dominated by Rayleigh scattering since there are no clouds, and the flux ratio is about four to five times smaller compared with the previous case and the overall flux ratio is on the order of 10$^{-10}$. This may thus be a limiting scenario that corresponds to the nominal precision achievable with exoplanet direct-imaging missions \citep[e.g.,][]{Seager2019, Roberge2018}. With the assumed S/N of 20, the full posterior distributions are reported in Appendix~\ref{sec:A_sky}. 
	
	The posterior distributions show that no clouds are detected and the abundance of gaseous H$_2$O should be low and negligible, as expected. The surface pressure and the surface albedo are therefore the most important parameters responsible for the baseline of the spectrum. The two parameters are therefore correlated with each other. The posterior distribution of the surface albedo, as well as the surface pressure, suggests a small bias from the true value. This bias could be linked with the moderate overestimation of the surface gravity (which is correlated with both surface albedo and pressure) and therefore the radius of the planet. On the atmospheric composition, the retrieval algorithm is able to tell that the atmosphere has a very high abundance of CO$_2$ and does not have a significant presence of other absorbing gasses, but also permits the interpretation of an N$_2$-dominated atmosphere. We consider this interpretation ``somewhat correct,'' as N$_2$ is a largely inactive gas.
	
	The abundance of O$_2$ in this atmosphere is moderately constrained, with a tail toward the low abundance. This is because the assumed abundance of O$_2$ is small, and thus the O$_2$ features in the spectrum are not as prominent as in the previous cases.
	 
	\begin{deluxetable}{cccc}
		\tablecaption{Atmospheric parameters used to simulate the dry CO$_2$-dominated atmosphere scenario and the retrieval results. The error bars of the retrieval results correspond to the 95\% confidence interval (i.e., 2$\sigma$). \label{tab:co2sky}}
		\tablehead{
			\colhead{Parameter} & \colhead{Input} & \colhead{Optical $+$ NIR} & \colhead{Optical only}}
		\startdata
		$Log(P_{0})$ [Pa] & $5.00$ & $5.71^{+0.26}_{-0.18}$ & $5.51^{+1.77}_{-0.78}$\\
		$Log(P_{top, H_2O})$ [Pa] & $-$ & $6.47^{+1.44}_{-2.49}$ & $4.10^{+3.57}_{-3.73}$\\
		$Log(D_{cld, H_2O})$ [Pa] & $-$ & $2.98^{+1.89}_{-2.47}$ & $4.95^{+3.29}_{-4.47}$\\
		$Log(CR_{H_2O})$ & $-$ & $-8.49^{+3.85}_{-3.09}$ & $-5.72^{+5.25}_{-5.77}$\\
		$Log(VMR_{H_2O})$ & $-$ & $-7.25^{+3.93}_{-3.42}$ & $-13.94^{+7.63}_{-9.90}$\\
		$Log(VMR_{CH_4})$ & $-5.96$ & $-7.54^{+1.73}_{-2.88}$ & $-13.68^{+7.08}_{-10.18}$\\
		$Log(VMR_{CO_2})$ & $-0.01$ & $-0.93^{+0.24}_{-0.23}$ & $-13.61^{+9.38}_{-9.74}$\\
		$Log(VMR_{O_2})$ & $-2.79$ & $-6.67^{+2.89}_{-3.70}$ & $-14.28^{+8.59}_{-9.48}$\\
		$Log(VMR_{O_3})$ & $-$ & $-9.34^{+1.22}_{-1.28}$ & $-14.36^{+5.95}_{-9.60}$\\
		$Log(VMR_{N_2})$ & $-1.55$ & $-0.05^{+0.02}_{-0.05}$ & $-0.01^{+0.01}_{-0.01}$\\
		$A_g$ & $0.2$ & $0.34^{+0.05}_{-0.04}$ & $0.30^{+0.48}_{-0.12}$\\
		$Log(g\ [cgs])$ & $2.99$ & $3.22^{+0.06}_{-0.05}$ & $3.17^{+0.37}_{-0.23}$\\
		$\mu$ & $43.55$ & $29.89^{+1.41}_{-0.77}$ & $28.01^{+0.01}_{-0.01}$\\
		\enddata
	\end{deluxetable}
	
	Without the NIR band (see Appendix~\ref{sec:A_sky}), however, the very high abundance of CO$_2$ cannot be retrieved. In this case, the retrieval algorithm mistakenly reports a 100\% N$_2$ atmosphere, without clouds or any other gases. Also, degenerate solutions are found regarding the surface pressure, the surface albedo, and the gravity of the planet, as these parameters are correlated with each other. We consider the spectral interpretation without the NIR band to be incorrect as it misses the main feature of the atmosphere, and this error is again due to the absorption features of gaseous CO$_2$ in the NIR band.
	
	\section{Discussion} \label{sec:discussion}
	
	\subsection{Comparison with previous studies}
	
	The results on the spectral characterization of a modern Earth analog presented in Section~\ref{sec:earth-like} can be compared with the results presented in \cite{Feng2018}. That paper also considered an Earth-like atmosphere and explored the effects on the atmospheric constraints with different S/N and spectral resolutions of a synthesized reflected spectrum in the wavelength range between $0.4-1.0\ \mu$m. That paper suggested that if a spectral resolution of 70 is considered, the minimum S/N required is 10 in order to characterize clouds and VMR of absorbing gases in the atmosphere. The required S/N bumps to 15 if a spectral resolution of 140 is instead assumed. 
	
	Our findings are in general agreement with \cite{Feng2018}. \exorelr\ presents the feature to not assume the dominant gas but let the spectrum speak for itself. With the inclusion of the NIR band ($1.0-1.8\ \mu$m), we find that an N$_2$-O$_2$ atmosphere is retrieved at the S/N and spectral resolution consistent with \cite{Feng2018}. Without the NIR band, the retrieval can still find the atmosphere to be O$_2$-dominated but will miss the N$_2$. These results show that a modern Earth analog is really an ideal case for spectral characterization in the reflected light, thanks to the richness of the spectral features (Fig.~\ref{fig:earth-like}). We suggest that S/N$=$5 is not enough to characterize the atmospheric composition and clouds in this case, and an S/N$>$10 and ideally 20 would be required.
	
	
	\subsection{The importance of the NIR band}
	
	A modern Earth analog is one of the many possibilities we might encounter when exploring terrestrial planets through direct imaging. Earth itself had an atmosphere very different from the present day for most of its past \citep[e.g.,][]{catling2020archean} and the other rocky planets in the Solar System have CO$_2$-dominated atmospheres, indicating the outcomes of rocky planet evolution are diverse. Therefore, the ability to measure the bulk atmospheric composition appears to be an inalienable part of the ability to characterize a terrestrial exoplanet. \exorelr\ was designed to quantify the requirement to measure the bulk atmospheric composition \citep{Damiano2021}. Here, based on the limited exploration of the scenarios that resemble Archean Earth and planets having CO$_2$-dominated atmospheres, we suggest that the optical band alone ($0.4-1.0\ \mu$m) is not sufficient to characterize terrestrial exoplanets found by direct imaging, and including the NIR band ($1.0-1.8\ \mu$m) can provide an adequate solution.
	
	    \begin{deluxetable*}{p{0.2\textwidth}|p{0.35\textwidth}|p{0.35\textwidth}}
		\tablecaption{The capability to characterize terrestrial planets with the reflected light spectroscopy in the optical and NIR bands. The high-level results are summarized from the retrievals on the four planetary scenarios with moderate spectral resolution and S/N. \label{tab:ret_summ}}
		\tablehead{
			\colhead{Scenario} & \colhead{Optical $+$ NIR} & \colhead{Optical only}}
		\startdata
		Modern Earth & \textbf{Correct:} N$_2$/O$_2$ atmosphere with H$_2$O, O$_3$ and CH$_4$ & \textbf{Somewhat Correct:} O$_2$ atmosphere with H$_2$O and O$_3$\\
		\hline
		Archean Earth & \textbf{Correct:} N$_2$ atmosphere with H$_2$O, CO$_2$, CH$_4$, and little O$_2$ & \textbf{Incorrect:} H$_2$O atmosphere with high clouds, CH$_4$ more than truth by $>1$ order of magnitude, and little CO$_2$\\
		\hline
		Habitable Venus (CO$_2$ atmosphere with O$_2$ and clouds) & \textbf{Correct:} CO$_2$ atmosphere with O$_2$, H$_2$O, O$_3$, and possibly N$_2$ & \textbf{Incorrect:} Massive N$_2$ atmosphere without clouds and with O$_2$, H$_2$O and O$_3$ less than truth by $>2$ orders of magnitude\\
		\hline
		Dry CO$_2$ atmosphere & \textbf{Somewhat Correct:} N$_2$/CO$_2$ atmosphere without clouds or H$_2$O & \textbf{Incorrect:} N$_2$ atmosphere without clouds, H$_2$O, or CO$_2$\\
		\enddata
	\end{deluxetable*}
	
	
	Table~\ref{tab:ret_summ} summarizes the main findings from the retrieval studies in Section~\ref{sec:result}, and compares what we would learn about the terrestrial planets with the optical band alone to that from both the optical and the NIR bands. Recall that our optical only cases assume a spectral resolution of 140 and an S/N of 20 in $0.4-1.0\ \mu$m, which is consistent with the best capability envisioned by existing mission studies for the optical band \citep[e.g.,][]{Roberge2018,Seager2019,Gaudi2020}. Even so, the retrievals misinterpret the optical-only spectra of all the scenarios that go beyond the modern Earth analog. Without the NIR band, the retrievals converge to solutions that are spectroscopically degenerate (Figs.~\ref{fig:earth-like}-\ref{fig:S1_c}), and having fundamentally different atmospheric composition from the truths (Table~\ref{tab:ret_summ}). It is striking that the modern Earth analog is the only scenario, among the four explored, for which the retrieval results from the optical-only spectrum can be deemed acceptable. In the other cases where CO$_2$ plays a more prominent role in shaping the spectra, the retrievals from the optical-only spectra return erroneous solutions.
	
	The specifics of the degenerate solutions may depend on the build-in assumptions of \exorelr, for example, the parameterization of the cloud density (Fig.~\ref{fig:profile}). However, our results indicate that at least one degenerate solution exists for each of the scenarios. Allowing more freedom in the cloud parameterization \citep[e.g.,][]{Feng2018,carrion2020directly} will result in potentially more degenerate solutions.
	
	The degenerate solutions found appear to be physical and thus unlikely to be ruled out by planetary models. The degenerate solution in the Archean Earth case resembles an H$_2$O steam atmosphere with high mixing ratios of CH$_4$, which by itself may be plausible from an atmospheric evolution standpoint \citep[e.g.,][]{zahnle1988evolution}. CH$_4$ should react with H$_2$O to produce CO$_2$, but the optical-only retrieval does not provide any constrain on the mixing ratio of CO$_2$, and the lifetime of CH$_4$ can be long if the atmosphere is massive and has some H$_2$ in the mixture \citep[e.g.,][]{Hu2014B2014ApJ...784...63H}. To maintain a steam atmosphere would require the planet to be sufficiently irradiated, but for a planet close to the inner edge of the habitable zone, it would be difficult to eliminate this degeneracy solely based on theoretical calculations of the planet's temperature, as those calculations have uncertainties and are sensitive to minor gases in the atmosphere \citep[e.g.,][]{way2020venusian,turbet2021day}. Similarly, the degenerate solution found in the CO$_2$-dominated atmosphere with clouds case involves a dry and massive atmosphere made of N$_2$, which would be uninhabitable and cannot be easily discounted from theoretical models (e.g., planetary and atmospheric evolution, thermal structure, and atmospheric chemistry).
	
	What if we let go of the objective of determining the bulk atmospheric composition and only focus on detecting the biosignature gases in these scenarios? Both O$_2$ and O$_3$ are detectable with the optical-only spectra, but their mixing ratios in a CO$_2$-dominated atmosphere would not be correctly retrieved unless the NIR band is included (Table~\ref{tab:ret_summ}). Several false-positive scenarios have been proposed for O$_2$ and O$_3$ as biosignature gases \citep[e.g.,][]{hu2012photochemistry,domagal2014abiotic,wordsworth2014abiotic,meadows2018exoplanet}, and it would be important to measure their mixing ratios correctly and interpret the mixing ratios in the context of planetary environments. For a planet like Archean Earth, the combination of CO$_2$ and CH$_4$ can be a biosignature \citep[e.g.,][]{schwieterman2018exoplanet}. However, this biosignature would not be detected from the optical-only spectrum. From the above, we see that an optical-only spectrum is probably also insufficient for detecting biosignature gases in terrestrial exoplanets.
    
    Taken together, the results shown in this paper indicate that spectroscopy in the optical band ($0.4-1.0\ \mu$m) is not sufficient to characterize the terrestrial exoplanets found by direct imaging, in either determining the state of the atmosphere or detecting the biosignature gases. As an initial step, our retrieval exploration suggests that including the NIR band ($1.0-1.8\ \mu$m) would eliminate the degenerate solutions and enable the proper measurements of the state of the atmosphere and the biosignature gases. Further extending the wavelength coverage to $\sim2-5\ \mu$m could result in better characterization and sensitivity to more biosignature gases (Martin et al. 2022, submitted). Extending the wavelength coverage into the UV may be an alternative solution \citep{lisman2019occulting}. We will assess these options in future studies.
    
	
	\subsection{Vertical profile of H$_2$O}
	
	It is also notable that the mixing ratio of H$_2$O below the cloud is retrieved correctly and precisely (typical $2\sigma$ uncertainty $\sim0.5$ dec) for all the scenarios with H$_2$O based on the optical and NIR spectra (Tables~\ref{tab:earthlike}-\ref{tab:co2clouds}). This ability to constrain the mixing ratio of H$_2$O below the cloud comes from the \exorelr\ feature that connects the cloud density to the drop off of the mixing ratio of H$_2$O, as shown in Fig.~\ref{fig:profile}. We see that the cloud position is also retrieved correctly in these scenarios, and this is because the water cloud is not fully reflective but has its spectral features in the NIR band. 
	
	The characterization of the cloud position and the mixing ratio of H$_2$O below the cloud will result in the constraints of the vertical profile of water vapor on a terrestrial exoplanet. This vertical profile can then be compared with climate model predictions to provide insight into the planet's hydrologic cycles and the existence of a large water reservoir on the surface.
	
	
	
	\section{Conclusions}
	\label{sec:conclusion}
	
	We use a Bayesian retrieval model \exorelr\ to study the spectral characterization capabilities of terrestrial exoplanets found by direct imaging. We consider plausible scenarios that range from planets similar to modern Earth and Archean Earth to planets having CO$_2$-dominated atmospheres, and perform spectral retrievals without any prior knowledge of the composition of the atmosphere. We find that a spectrum in $0.4-1.8\ \mu$m with moderate resolution ($R=140$ in the optical and $40$ in the NIR band) and an S/N$=$20 can distinguish each of these scenarios, determine the main atmospheric component (N$_2$, O$_2$, or CO$_2$), measure the mixing ratios of trace gases including H$_2$O, O$_3$, and CH$_4$, and constrain the pressure level of a cloud layer. These constraints amount to a comprehensive understanding of the state of the atmosphere, detection of O$_2$/O$_3$ or CO$_2$-CH$_4$ biosignature gases, and potential inference of liquid-water oceans on the surface.
	
	To achieve this vision of terrestrial exoplanet characterization will require spectroscopy that goes beyond the optical band of $0.4-1.0\ \mu$m. Consistent with \citet[][]{Feng2018}, we show that the optical-band spectroscopy at the resolution of at least $70-140$ and the S/N of at least $10-20$ can provide reasonable constraints on an Earth-like atmosphere. But for planets like Archean Earth or having CO$_2$-dominated atmospheres, the optical-band spectroscopy will be insufficient and suffer from physically plausible degenerate solutions. Including the NIR band will eliminate the degeneracies and prevent incorrect interpretation.
	
	It is thus crucial to consider a broad wavelength coverage that goes beyond the optical band when designing the future space telescope capable of terrestrial exoplanet imaging. This is, in a general sense, consistent with the proposed mission concepts HabEx \citep{Gaudi2020} and LUVOIR \citep{Roberge2018} that would provide the spectroscopic capability in the NIR band. Regardless of the technologies used for starlight suppression (i.e., coronagraph and starshade), it should be considered a high priority to obtain the broad-wavelength spectroscopy for as many targets as possible when evaluating mission architectures. The minimum requirement for spectral characterization of potentially habitable exoplanets should include, at least a fraction of, the NIR band. Future studies will further pinpoint the short and long wavelength cutoffs for an adequate spectrum of directly imaged terrestrial exoplanets.
	
	
	
	
	\section*{Acknowledgments}
	We thank Charles Lawrence, Bertrand Mennesson, Keith Warfield, and Rhonda Morgan for helpful discussion. This work was supported in part by the NASA WFIRST Science Investigation Teams grant \#NNN16D016T and Exoplanets Research Program grant \#80NM0018F0612. This research was carried out at the Jet Propulsion Laboratory, California Institute of Technology, under a contract with the National Aeronautics and Space Administration. 
	
	{	\small
		\bibliographystyle{apj}
		\bibliography{bib.bib}
	}
	
	\appendix
	
	\section{Scenario 1: An Earth-like Atmosphere} \label{sec:A_earth}
	
\begin{figure*}[!h]
		\centering
		\includegraphics[scale=0.425]{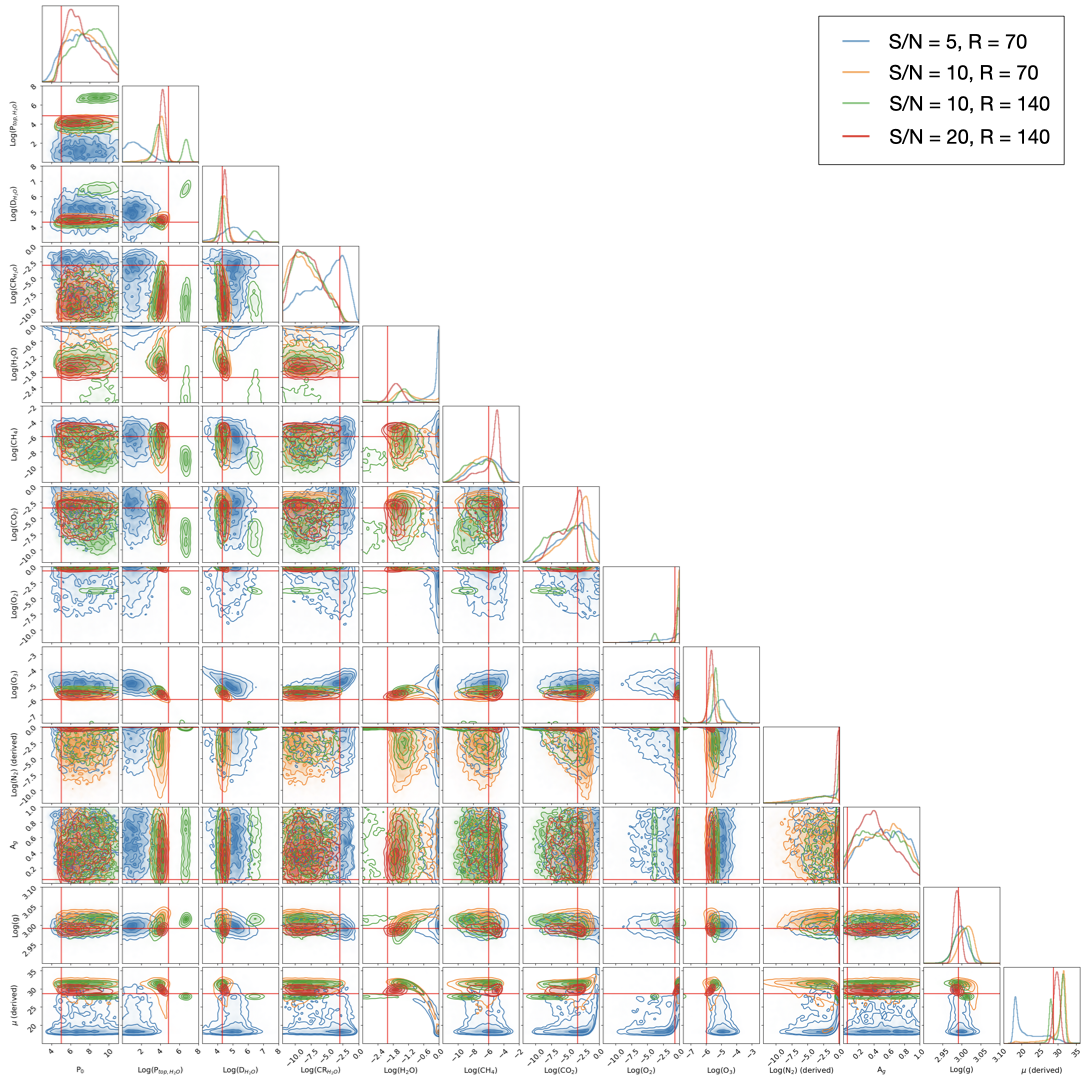}
		\caption{The full posterior distributions (corner plots) of the Earth-like scenario. The different cases explore a combination of different optical-band spectral resolution and S/N. The red lines in the corner plots refer to the true value used to simulate the data. \label{fig:post_earth_sign}}
	\end{figure*}

\begin{figure*}[!h]
		\centering
		\includegraphics[scale=0.425]{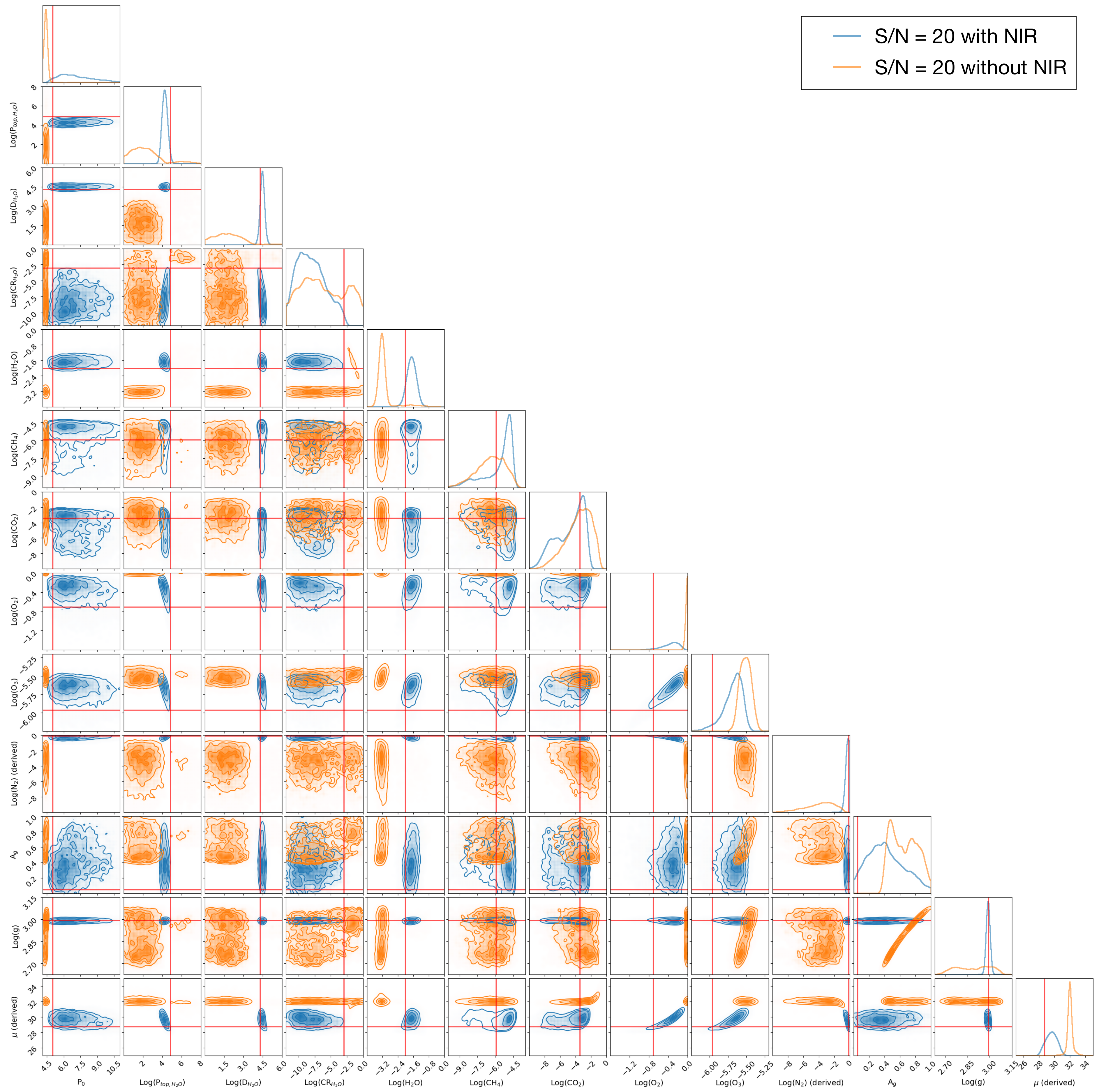}
		\caption{The full posterior distributions of the Earth-like scenario with (blue lines) and without (orange lines) the NIR wavelength band in 1.0 -- 1.8 $\mu$m. The red lines in the corner plots refer to the true value used to simulate the data. \label{fig:post_earth_ir}}
	\end{figure*}

	We simulate the reflected spectrum of an N$_2$-dominated atmosphere with O$_2$ being the second most abundant gas (Fig.~\ref{fig:earth-like}). We also include H$_2$O, CH$_4$, CO$_2$, and O$_3$ as minor absorbing gases. We also include a water cloud layer and a surface with albedo of 0.05. We use the synthesized data as input for \exorelr, and Fig.~\ref{fig:post_earth_sign} comprises the resulted posterior distributions of all the S/N and spectral resolution combinations discussed in Sec.~\ref{sec:scenarios}. Also, we run \exorelr\ on only the optical part of the spectrum to explore the benefit of having the NIR wavelength band in addition to the optical band. The result is shown in Fig.~\ref{fig:post_earth_ir}.
	
	\newpage
	
	\newpage
	
	\section{Scenario 2: An Archean Earth-like atmosphere} \label{sec:A_archean}
	
\begin{figure*}[!h]
		\centering
		\includegraphics[scale=0.425]{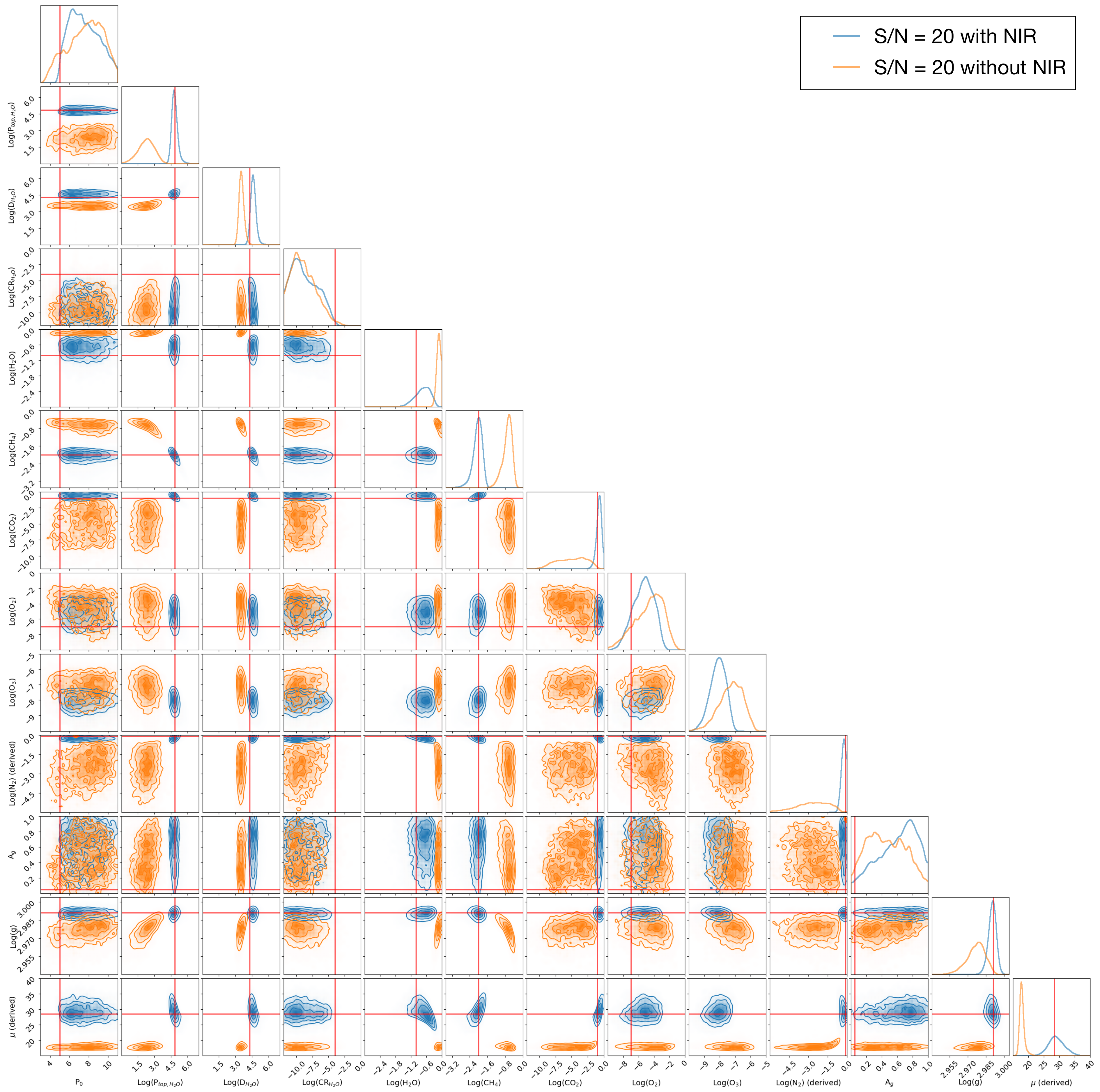}
		\caption{The full posterior distributions of the Archean Earth-like scenario with (blue lines) and without (orange lines) the NIR wavelength band. The red lines in the corner plots refer to the true value used to simulate the data. \label{fig:post_archean}}
	\end{figure*}

	Earth's atmosphere has experienced multiple evolutionary stages before reaching the current state. One of the main stages early in the history is called Archean. During this stage, the atmosphere had minimal O$_2$ and a higher amount of CO$_2$ and CH$_4$ with respect to the contemporary atmosphere. Therefore, we also simulate the reflected spectrum of a planet like the Archean Earth, assuming an N$_2$-dominated atmosphere with H$_2$O, CH$_4$, CO$_2$, and O$_2$ as minor absorbing gases (Fig.~\ref{fig:S1_e}). We still include a water cloud layer and a surface with albedo of 0.05. The synthesized data are used as input for \exorelr. We perform the retrieval with and without the IR part of the spectrum to explore the benefit of having the NIR wavelength band in addition to the optical. The result is shown in Fig.~\ref{fig:post_archean}.
	
	\newpage
	
	\section{Scenario 3: A CO$_2$-dominated atmosphere with O$_2$ and clouds} \label{sec:A_co2}
	
\begin{figure*}[!h]
		\centering
		\includegraphics[scale=0.425]{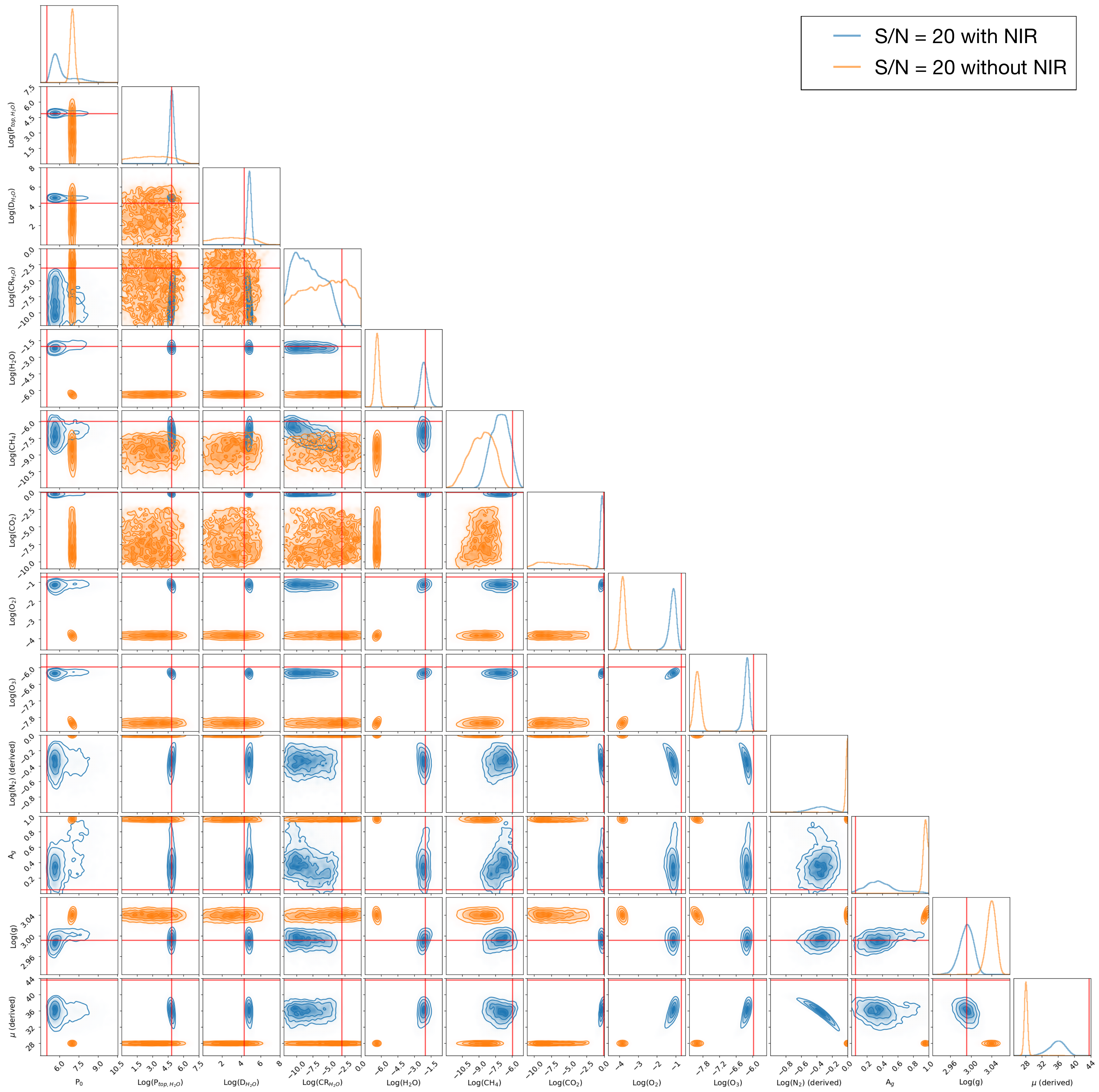}
		\caption{The full posterior distribution of the CO$_2$-dominated atmosphere with O$_2$ and clouds scenario with (blue lines) and without (orange lines) the NIR wavelength band. The red lines in the corner plots refer to the true value used to simulate the data. \label{fig:post_co2}}
	\end{figure*}

	We simulate the reflected spectrum of a CO$_2$-dominated atmosphere with O$_2$ and a cloud layer (Fig.~\ref{fig:S1_d}). We include H$_2$O, CH$_4$, and O$_3$ as minor absorbing gases. We also include a surface with albedo of 0.05. We run \exorelr\ on only the optical part of the spectrum to explore the benefit of having the NIR wavelength band in addiction to the optical. The result is shown in Fig.~\ref{fig:post_co2}.
	
	\newpage
	
	\section{Scenario 4: A dry CO$_2$-dominated atmosphere} \label{sec:A_sky}
	
\begin{figure*}[!h]
		\centering
		\includegraphics[scale=0.425]{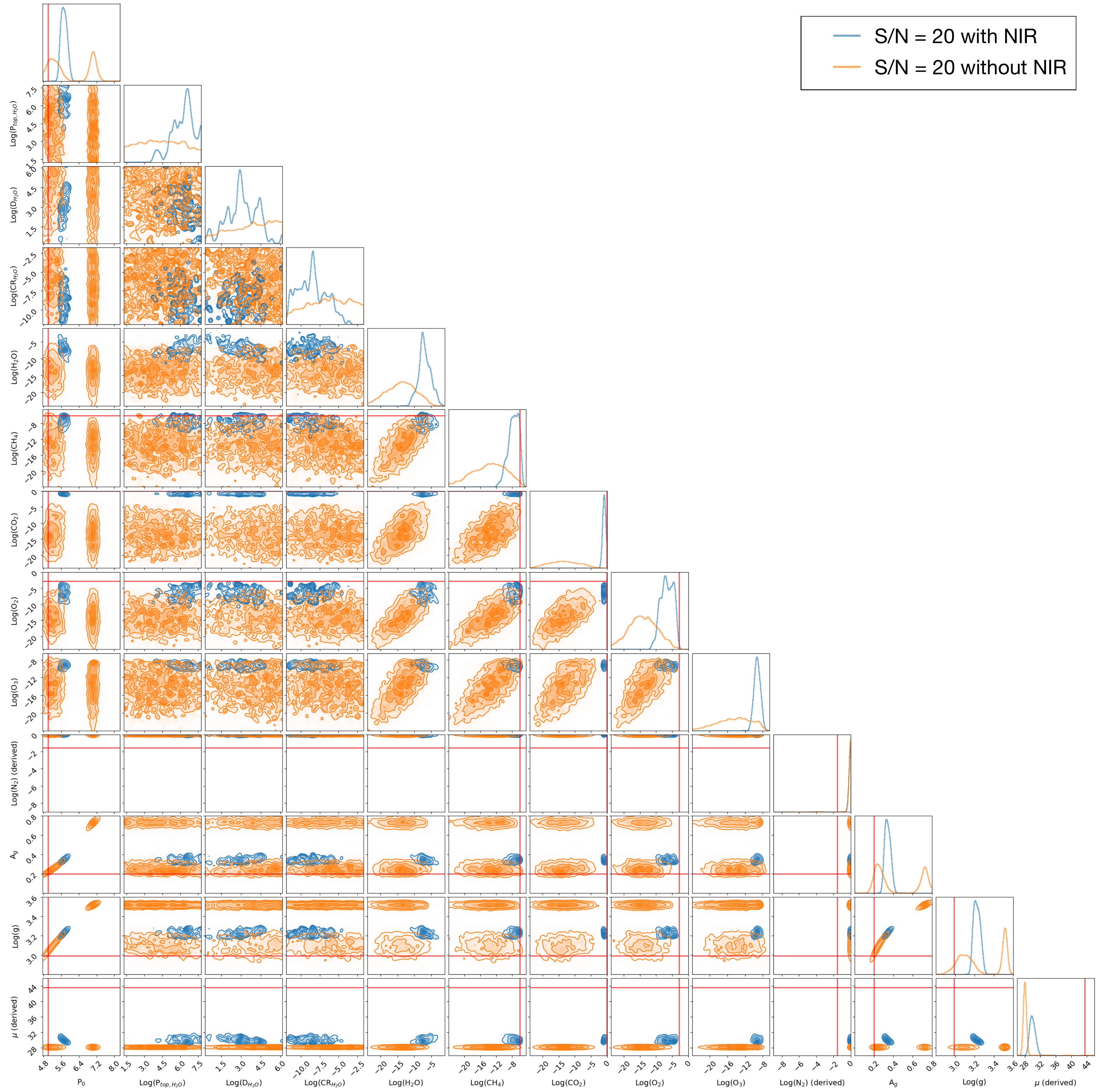}
		\caption{The full posterior distribution of the dry CO$_2$-dominated atmosphere without clouds scenario with (blue lines) and without (orange lines) the NIR wavelength band. The red lines in the corner plots refer to the true value used to simulate the data. \label{fig:post_marslike}}
	\end{figure*}

	We simulate the reflected spectrum of a dry CO$_2$-dominated atmosphere without clouds (Fig.~\ref{fig:S1_c}). We use a CO$_2$-dominated atmosphere with CH$_4$ and O$_2$ as minor absorbing gases. We include a surface with albedo of 0.2. We run \exorelr\ on only the optical part of the spectrum to explore the benefit of having the NIR wavelength band in addiction to the optical. We use the synthesized data as input for \exorelr, and the result is shown in Fig.~\ref{fig:post_marslike}.
	
	\newpage
	
\end{document}